\documentclass[
  aps,
  prl,
  twocolumn,
  amsfonts,
  amssymb,
  amsmath,
  superscriptaddress,
  floatfix
]{revtex4-1}

\usepackage[utf8]{inputenc}
\usepackage{bm}  
\usepackage{siunitx}
\usepackage{booktabs}
\usepackage{graphicx}
\usepackage[svgnames]{xcolor}
\usepackage{adjustbox}
\usepackage{tabularx}
\usepackage{xspace}  
\usepackage{comment}
\usepackage{tikz} 
\usepackage{import} 

\usepackage[hidelinks]{hyperref}

\usetikzlibrary{calc,tikzmark} 
\newlength{\imageheight}
\newcommand{\FontLabel}{\sffamily\bfseries\large}

\newcommand{\InsertLabels}[2]{%
  \begin{tikzpicture}[overlay,remember picture]
        \node[font=\FontLabel] at ( $ (pic cs:a) +(#1,\imageheight-#2) $ ){a};%
        \node[font=\FontLabel] at ( $ (pic cs:b) +(#1,\imageheight-#2) $ ){b};%
        \node[font=\FontLabel] at ( $ (pic cs:c) +(#1,\imageheight-#2) $ ){c};%
        \node[font=\FontLabel] at ( $ (pic cs:d) +(#1,\imageheight-#2) $ ){d};%
    \end{tikzpicture}   
}

\newcommand{\InsertLabelsss}[2]{%
  \begin{tikzpicture}[overlay,remember picture]
        \node[font=\FontLabel] at ( $ (pic cs:a) +(#1,\imageheight-#2) $ ){a};%
        \node[font=\FontLabel] at ( $ (pic cs:b) +(#1,\imageheight-#2) $ ){b};%
    \end{tikzpicture}   
}
\newcommand*{\vb}[1]{\boldsymbol{#1}}  
\newcommand*{\dd}{\mathrm{d}}  

\newcommand*{\gradient}{\vb{\nabla}}

\newcommand*{\vvec}{\vb{v}}

\newcommand*{\xvec}{\vb{x}}

\newcommand*{\lvec}{\vb{l}}
\newcommand*{\rvec}{\vb{r}}

\newcommand*{\vrms}{v_{\mathrm{rms}}}

\newcommand*{\vortvec}{\vb{\omega}}
\newcommand*{\Cloop}{\mathcal{C}}

\newcommand*{\zetainv}{\zeta^{\mathrm{IEC}}}
\newcommand*{\zetadir}{\zeta^{\mathrm{DEC}}}
\newcommand*{\lambdainv}{\lambda^{\mathrm{IEC}}}
\newcommand*{\lambdadir}{\lambda^{\mathrm{DEC}}}

\newcommand*{\mean}[1]{\langle #1 \rangle}
\newcommand*{\Lint}{L_\text{I}}  
\newcommand*{\kint}{k_\text{I}}

\begin{document}

\title{Exploring the Equivalence between two-dimensional Classical and Quantum Turbulence through Velocity Circulation Statistics}

\author{Nicol\'as P. M\"uller}
\affiliation{%
Université Côte d'Azur, Observatoire de la Côte d'Azur, CNRS,Laboratoire Lagrange, Boulevard de l'Observatoire CS 34229 - F 06304 NICE Cedex 4, France
}
\affiliation{
Laboratoire de Physique de l'École normale supérieure, ENS, Université PSL, CNRS, Sorbonne Université, Université de Paris, Paris, France
}

\author{Giorgio Krstulovic}
\affiliation{%
Université Côte d'Azur, Observatoire de la Côte d'Azur, CNRS,Laboratoire Lagrange, Boulevard de l'Observatoire CS 34229 - F 06304 NICE Cedex 4, France
}
\date{\today}

\begin{abstract}
  We study the statistics of velocity circulation in two-dimensional classical and quantum turbulence. We perform numerical simulations of the incompressible Navier--Stokes and the Gross--Pitaevskii (GP) equations for the direct and inverse cascades. Our GP simulations display clear energy spectra compatible with the double cascade theory of two-dimensional classical turbulence. 
  In the inverse cascade, we found that circulation intermittency in quantum turbulence is the same as in classical turbulence. We compare GP data to Navier--Stokes simulations and experimental data from [Zhu~et~al.~{\it Phys.~Rev.~Lett.}~{\bf130},~214001(2023)].
  In the direct cascade, for nearly-incompressible GP-flows, classical and quantum turbulence circulation displays the same self-similar scaling. When compressibilty becomes important, quasi-shocks generate quantum vortices and the equivalence of quantum and classical turbulence only holds for low-order moments.
  Our results establish the boundaries of the equivalence between two-dimensional classical and quantum turbulence.
\end{abstract}

\maketitle

The chaotic spatiotemporal motion of turbulent flows is a complex multi-scale phenomenon occurring in a wide variety of systems in nature \cite{Frisch1995,Pope2000,Batchelor2000}. One of the most fascinating properties of three-dimensional (3D) turbulence is that energy is transferred from large to small structures at a constant energy rate, in a process known as direct energy cascade. 
Some geophysical flows, like atmospheres or oceans, present a quasi-two-dimensional (2D) behavior due to the suppression of motion in one direction induced by rotation or stratification \cite{Davidson2013,Young2017}. 
Contrary to the 3D case, two-dimensional (2D) turbulence exhibits an inverse energy cascade (IEC), in which energy is transferred towards large scales leading to the formation of large-scale coherent structures \cite{Tabeling2002,Boffetta2012}. Moreover, enstrophy $\Omega$ --- defined as one-half the mean-squared vorticity $\Omega = \langle \omega^2 \rangle / 2$ --- is transferred towards smaller scales in a process known as direct enstrophy cascade (DEC) \cite{Kraichnan1967,Leith1968,Batchelor1969}. 

Turbulence also takes place in superfluids, such as $^4$He and Bose-Einstein condensates (BEC) \cite{Pethick2008,Rousset2014,Tsatsos2016}. Due to quantum mechanics, low-temperature superfluids are characterized by the complete absence of viscous effects. In 2D quantum fluids, vorticity is concentrated in topological point-like defects with a quantized circulation. The mutual interaction of these structures, known as quantum vortices, leads to the out-of-equilibrium state known as quantum turbulence (QT) \cite{Barenghi2014}. Experiments in 2D BECs and quantum fluids of exciton-polaritons have shown evidence of an IEC through the formation of Onsager vortex clusters \cite{Johnstone2019,Gauthier2019,Panico2023}.
Direct numerical simulations (DNS) of 2D and quasi-2D quantum turbulence have shown the development of an IEC with the presence of a Kolmogorov energy spectrum \cite{Bradley2012, Reeves2013, Muller2020a}. 
The vorticity field in quantum fluids is a superposition $\delta$-Dirac supported terms, making enstrophy ill-defined mathematically. Still, it can be phenomenologically related to the total number of vortices, which in general can decrease due to vortex-antivortex annihilation \cite{Numasato2010a}. However, if compressible effects are neglected, the number of vortices will be bounded by its initial value and remain almost constant. In this case, one can expect the development of an enstrophy cascade \cite{Reeves2017}.

Another very interesting property of 2D turbulence is the lack of intermittency in the IEC. 
Velocity increments $\delta \vvec_r = \vvec(\xvec+\rvec) - \vvec(\xvec)$ at a length scale $r$ in 2D turbulent flows follow close-to-Gaussian statistics \cite{Boffetta2000, Paret1998}, in stark contrast with 3D turbulence where velocity fluctuations are strong \cite{Frisch1995,Iyer2020}. 
As a consequence, the structure functions of order $p$ defined as $S_p = \langle \delta v_r^p \rangle$ follow a self-similar scaling within the inertial range $S_p \sim r^{\zeta_p}$ with $\zetainv_p = p/3$. 
The DEC is also non-intermittent as the velocity field in this regime is smooth, and the scaling exponents follow $\zetadir_p = p$ \cite{Paret1999}. 

\begin{figure*}
  \centering
  \includegraphics[width=1\textwidth]{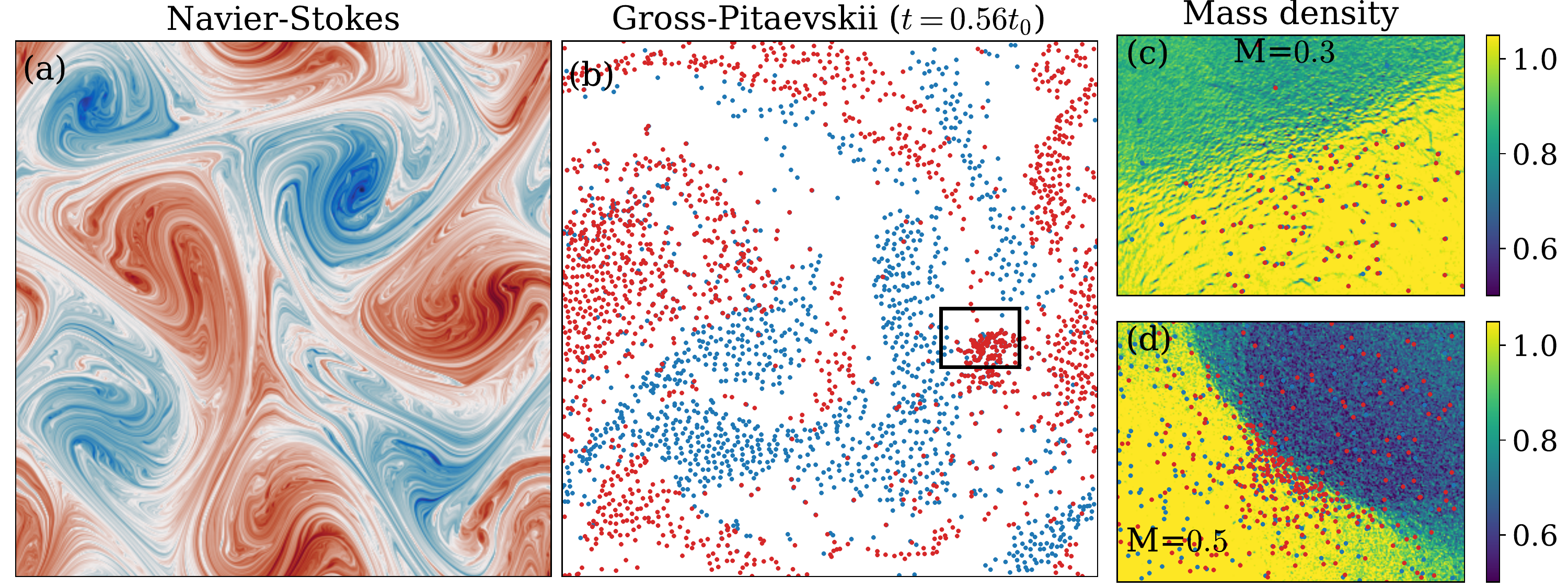}
  \caption[]{%
    Visualization of vorticity in (a) classical and (b) quantum turbulence in the enstrophy cascade. For Navier--Stokes, we show the vorticity field $\omega(x,y)$ (RUN NS-dir). For Gross--Pitaevskii we show the sign and position of individual vortices at $t=0.56t_0$ with $t_0 = L_0 / \vrms$ (RUN GP-dir-M03). Panel (c) shows the density field $|\psi|^2$ exhibiting a mild quasi-shock (area indicated with a rectangle in (b)) for a flow with a Mach number $M=0.3$, while (d) corresponds to a quasi-shock for a flow with $M=0.5$ (RUN GP-dir-M05). 
    Full movies of the GP evolution are provided in the Supplemental Material.}
    \label{fig:visualization}
\end{figure*}

An alternative way of studying turbulence intermittency is through the velocity circulation around an area $A$ enclosed by a loop $\Cloop$, defined as $\Gamma = \oint_\Cloop \vvec \cdot \dd \lvec$. 
High-resolution DNS of 3D classical turbulence (CT) have shown that circulation moments in the inertial range are less intermittent than velocity increments when compared with the self-similar Kolmogorov prediction \cite{Iyer2019, Frisch1995, Iyer2020, Iyer2021, Migdal1994}. 
Recent experimental studies in quasi-2D CT showed that circulation in the DEC is non-intermittent, while in the IEC, it surprisingly presents anomalous deviations \cite{Zhu2023}.
The study of circulation in QT turns out to be very convenient due to the discrete nature of quantum vortices. 
Indeed, DNS and experiments of 3D QT have shown that circulation statistics is very similar to 3D CT \cite{Muller2021, Polanco2021, Muller2022a}. This result implies that the nature of circulation at small scales becomes irrelevant in the inertial scales and motivates the use of quantum fluids and circulation statistics as a discrete system to understand intermittency in CT. 

In this Letter, we compare the statistics of velocity circulation in two-dimensional quantum and classical turbulence, both in the inverse and direct cascades. Using DNS, we characterize the intermittent behavior of these two regimes, finding differences and similarities between 2D CT and QT. 

The dynamics of an incompressible two-dimensional classical fluid is described by the NS equation, which in terms of the vorticity field $\omega(\bm{r}, t) = -\nabla^2 \phi$ is written as
\begin{equation}
  \partial_t \omega + \left\{\omega, \phi\right\} = \nu \nabla^2 \omega - \alpha \omega + f
  \label{eq:ns}
\end{equation}
with $\phi$ the stream function such that the velocity field is $(u, v) = (\partial_y \phi, -\partial_x \phi)$, the Poisson brackets are defined as $\left\{\omega, \phi\right\} = \partial_x \omega \partial_y \phi - \partial_y \omega \partial_x \phi$, $\nu$ is the kinematic viscosity, $\alpha$ is a linear friction preventing the formation of a large-scale condensate, and $f$ an external forcing. 
The dynamics of a quantum fluid composed of weakly interacting bosons at zero temperature is described by the GP equation
\begin{equation}
  i \partial_t \psi = \frac{c}{\sqrt{2}\xi} \left( -\xi^2 \nabla^2 \psi + \frac{|\psi|^2}{n_0} \psi - \psi\right)
  \label{eq:gp}
\end{equation}
where $\psi$ is the condensate wave function, $n_0$ is the ground state particles density, $c = \sqrt{g n_0/m}$ the speed of sound and $\xi = \hbar / \sqrt{2mgn_0}$ the healing length, which is proportional to the quantum vortex core size. Here, $m$ is the mass of the bosons, and $g$ is the coupling constant. 
It is important to notice that in the NS equation, circulation takes real values while, in the GP equation, it is discrete as $\Gamma = n\kappa$, with $n\in \mathbb{Z}$ the vortex charge and $\kappa=h/m=2\pi\sqrt{2}c\xi$ the quantum of circulation. 

Equations~\eqref{eq:ns} and \eqref{eq:gp} are solved using a standard pseudospectral method in a periodic two-dimensional domain. We use a Runge-Kutta temporal scheme of order $2$ for NS and order $4$ for GP. For each equation, we optimize parameters to achieve the largest possible scale separation for each cascade. For NS, we use $6144^2$ grid points and $8192^2$ for GP. To generate the IEC in NS, we force at small scales and dissipate by the friction term and by viscous dissipation. For the DEC, forcing is applied at large scales and no friction is included. For both cascades, we average several hundred fields from the stationary state. For the GP equation, the total energy is conserved, but incompressible energy (vortices) is irreversibly converted into sound. Therefore, GP simulations can be seen as decaying turbulent runs. We analyze data when turbulence is the strongest. For both cascades, we generate an ensemble of initial conditions with most of their energy concentrated at a target wave number. These flows are obtained by a minimization method that reduces the acoustic contribution  \cite{Nore1997,Muller2020a}. 
As we intend to compare QT with incompressible CT, GP reference runs have a small initial Mach number ${\rm M}=\vrms/c \leq 0.3$ where compressible effects are negligible. For comparison, we also prepare GP DEC initial data with ${\rm M}=0.5$. See Supplemental Material (SM) for details on parameter values and initial conditions. 
Relevant length scales in the turbulent regimes are shown in Table~\ref{tab:runs}.
\begin{table}
  \begin{tabular*}{\linewidth}{@{\extracolsep{\fill}} l c c c c c }
    \hline \hline
    RUN & N & $\Lint / L_0$  & $L_0 / \eta$ & $\ell / L_0$ & $L_0 / \xi$  \\
    \hline
    NS-inv & 6144 & 176   &  -     & - & -   \\
    NS-dir & 6144 & 0.65  &  3788  & - & -   \\
    GP-inv & 8192 & 16.0  &  -     & 0.75  & 45.51  \\
    GP-dir-M03 & 8192 & 1.7   &  -     & 0.036 & 2731 \\
    GP-dir-M05 & 8192 & 1.55  &  -     & 0.013 & 4096 \\
    \hline \hline
  \end{tabular*}
\caption{ Typical length scales of numerical simulations of the NS and GP equations, with $N$ the linear collocation points. $L_0$ corresponds to the forcing scale $L_f$ in NS, and the initial condition characteristic length scale $L_{\mathrm{IC}}$ in GP. $\Lint$ is the integral length scale, $\eta$ the Kolmogorov length scale, $\ell$ the inter-vortex distance and $\xi$ the healing length. Runs GP-dir-M03 and GP-dir-M05 are both forced at large scales, but the initial flows have different Mach numbers $M=0.3$ and $M=0.5$, respectively.   
}
\label{tab:runs}
\end{table}
Figure \ref{fig:visualization} shows the vorticity field in two-dimensional classical and quantum flows in the DEC regime. Both systems display the typical large-scale thin elongated structures of the enstrophy cascade, despite the fundamental small-scale difference of vortices. 
Such structures can create strong density gradients eventually leading to quasi-shocks and the spontaneous generation of vortices (Fig.~\ref{fig:visualization}.c-d). As it will be seen later, for low-Mach flows (Fig.~\ref{fig:visualization}.b-c) those events are weak and scarce enough in time and space to not influence turbulence statistics (see movies in the SM).

\begin{figure}
  \centering
  \includegraphics[width=.48\textwidth]{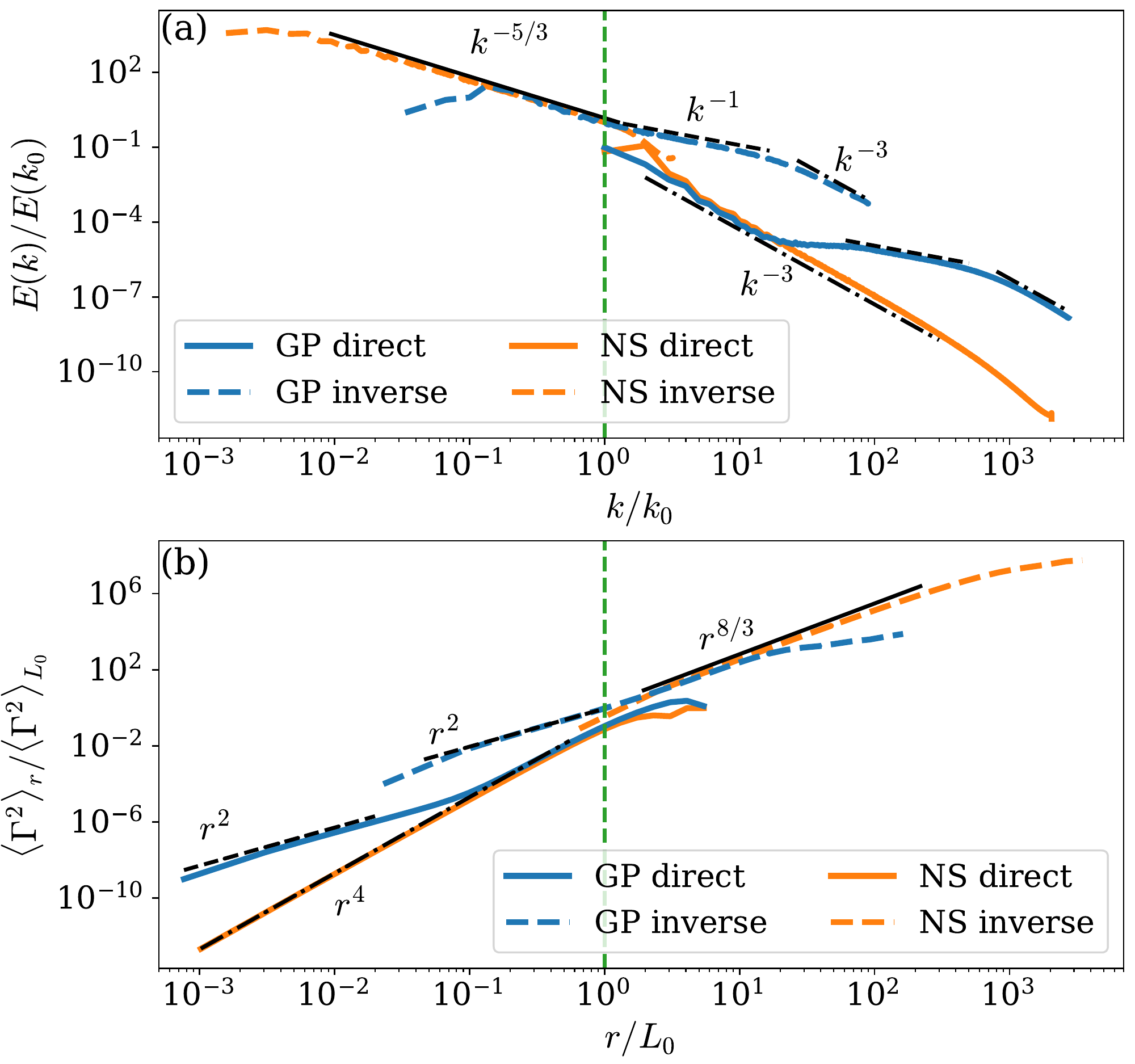}
  \caption[]{%
    (a) Incompressible energy spectra and (b) circulation variance in NS and GP for both the IEC and DEC runs, with Mach number $M=0.3$. $L_0$ indicates the forcing scale in NS and the initial injection length scale in GP. Direct cascade curves are vertically shifted for better visualization. Other characteristic length scales are reported in Table~\ref{tab:runs}.  
    }
    \label{fig:spectra_and_variance}
\end{figure}

According to the Kraichnan-Leith-Batchelor (KLB) theory \cite{Kraichnan1967,Leith1968,Batchelor1969}, the energy spectra in the inverse and direct cascade regimes, neglecting logarithmic corrections, follow
\begin{eqnarray}
  E(k) &=& C_E \epsilon^{2/3} k^{-5/3} \quad \mathrm{for}\quad \kint<k<k_f \\
  E(k) &=& C_\Omega \beta^{2/3} k^{-3} \quad \mathrm{for}\quad k_f<k<k_\eta,\label{Eq:KLB}
\end{eqnarray}
where $\epsilon$ and $\beta$ are the energy and enstrophy dissipation rates, respectively, and $C_E$ and $C_\Omega$ are dimensionless universal constants. 
The inertial range for the IEC lays between the integral scale wave number $\kint = 2\pi / \Lint$, with $\Lint = 2\pi \int k^{-1}E(k) \dd k / \int E(k) \dd k$, and the forcing wave number $k_f$. The DEC takes place between the forcing and the dissipation wave numbers $k_\eta$, with $\eta$ the enstrophy dissipation length scale $\eta = \nu^{1/2} / \beta^{1/6}$. 
Figure~\ref{fig:spectra_and_variance}(a) shows the incompressible energy spectra of all four simulations. 
The subscript $_0$ denotes the forcing scale for NS or the initial condition scale for GP. For small wave numbers $k/k_0<1$, we observe the $k^{-5/3}$ scaling law of the IEC in both, classical and quantum 2D turbulence. 
For large wave numbers $k/k_0 > 1$ the energy spectra exhibit a $k^{-3}$ scaling law corresponding to the DEC, which in quantum turbulence takes place  
between $k_0<k<k_\ell$, with $\ell=2\pi/k_\ell$ the inter-vortex distance. 
We verified that within the inertial range of the IEC the energy flux becomes close to constant taking negative values, while the enstrophy flux in the DEC becomes positive (see SM).
In the GP case, we also observe the development of two other scaling laws. Between the inter-vortex distance $\ell$ and healing length $\xi$ ($k_\ell$ and $k_\xi$ wave numbers, respectively), the dynamics is governed by single quantum vortices having an azimuthal velocity field $v(r) = \kappa/(2\pi r)$, which leads to a $k^{-1}$ energy spectrum \cite{Krstulovic2010,Bradley2012}. Note that in 3D QT, this is the range of scales in which Kelvin waves are observed \cite{Krstulovic2012,ClarkdiLeoni2017,Muller2020}. For $k>k_\xi$, there is a $k^{-3}$ scaling law due to the core of quantum vortices \cite{Krstulovic2010}. 

We now focus on the statistics of velocity circulation in 2D classical and quantum turbulence.
We compute the circulation $\Gamma_r = \oint_{\Cloop_r} \vvec \cdot \dd \lvec$ around squared planar loops of linear size $r$. Integrals are performed in Fourier space to take advantage of the spectral accuracy of the simulations \cite{Circulation}. For the GP equation, we obtain the velocity field from the condensate wave function as $\vvec = -\sqrt{2}c\xi \mathrm{Im}(\psi \gradient \psi^*) / \rho$, after performing a Fourier interpolation of $\psi$ to a resolution $32678^2$ to better resolve the vortex density profiles \cite{Muller2021}. 
For small scales $r/L_0 < 1$, the circulation variance $\mean{\Gamma_r^2}$ in CT follows the $r^4$ scaling expected for a smooth field that extends for the DEC and the diffusive scales (see Fig.~\ref{fig:spectra_and_variance}.b). In QT, it follows the $r^4$ scaling for $\ell < r < L_0$, and there is a second $r^2$ scaling given by the probability of finding a quantum vortex inside a loop for $r<\ell$ \cite{Muller2021}. The IEC inertial range takes place at large scales $L_0<r<\Lint$, where the circulation variance follows a $r^{8/3}$ scaling consistent with KLB theory. 

\begin{figure}
  \centering
  \includegraphics[width=.48\textwidth]{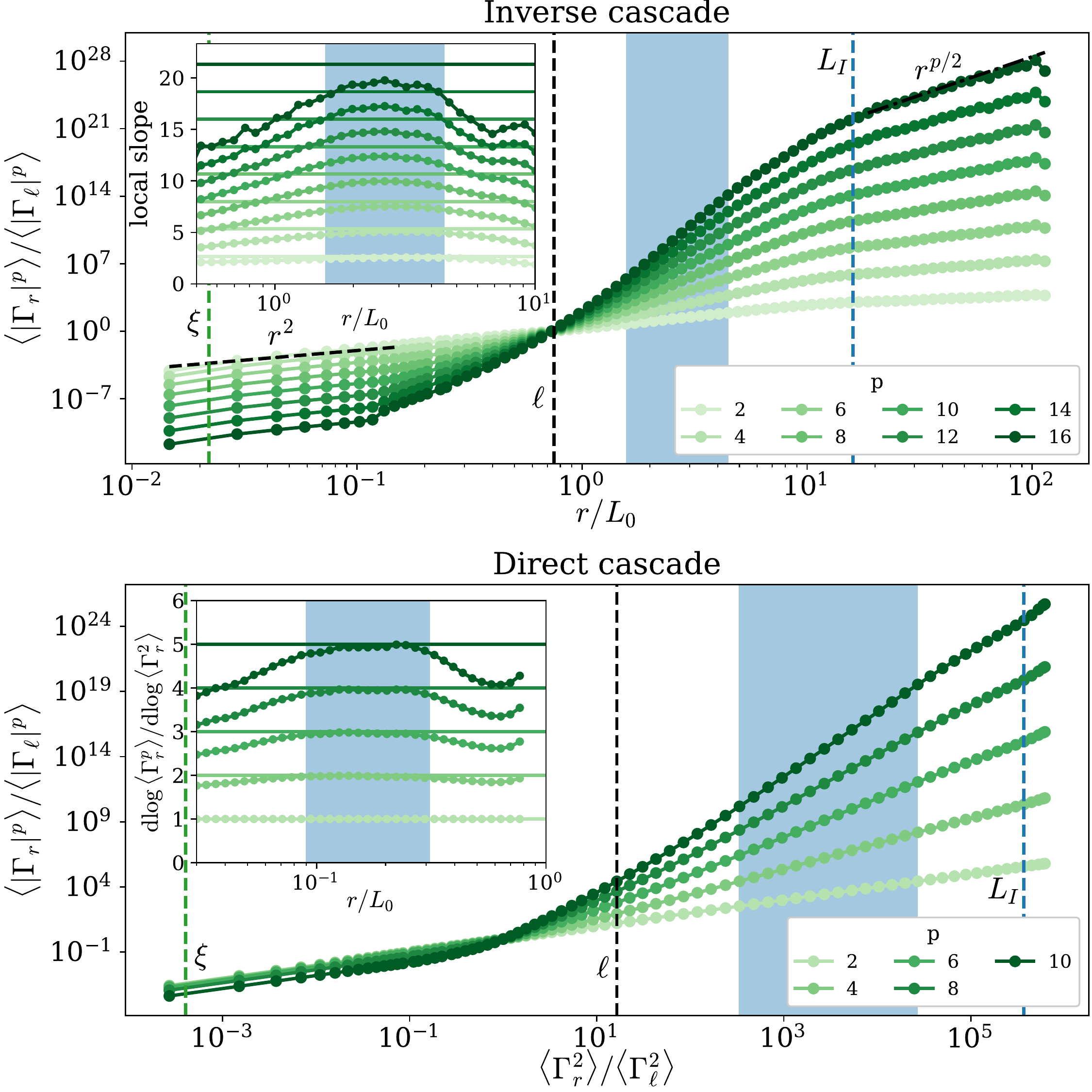}
  \caption[]{%
    Circulation moments in two-dimensional quantum turbulence for (a) the inverse energy cascade as a function of $r/L_0$ and (b) the direct enstrophy cascade as a function of the circulation variance. The insets display the local slopes defined as $\dd \log \langle |\Gamma_r|^p \rangle / \dd \log x$, with $x=r$ or $x=\mean{\Gamma_r^2}$. 
    }
    \label{fig:circulation_moments}
\end{figure}

To characterize the intermittent behavior of the two cascades, we compute the circulation moments $\langle |\Gamma_r|^p \rangle$ up to order $p=16$ in QT (see Fig.~\ref{fig:circulation_moments}) and CT (see SM). 
The good statistical convergence of high-order moments is shown in the SM. For the IEC in the inertial range $L_0<r<\Lint$, circulation moments display scaling laws that deviate from the self-similar prediction of $\lambda_p^{\mathrm{IEC}} = 4p/3$, obtained by dimensional arguments. This behavior is better observed in the local slopes displayed in the insets, defined as the logarithmic derivatives $\dd \log \langle |\Gamma_r|^p \rangle / \dd \log r$, which become flat in the inertial range. For the largest scales of the system $r>\Lint$, circulation moments follow a scaling $r^{p/2}$, which is smaller than the scaling of a system of randomly distributed vortices \cite{Polanco2021}. Such an exponent suggests an anti-correlation between vortices that could be induced by a gas of vortex dipoles. The behavior in this range of scales might also depend on the initial conditions and is likely to be non-universal. Further studies of this regime are left for future work.
In the DEC, we plot the extended-self-similar (ESS) moments with respect to the circulation variance and obtain the self-similar scaling $\lambda_p^{\mathrm{DEC}} = 2p$.

\begin{figure}
  \centering
  \includegraphics[width=.48\textwidth]{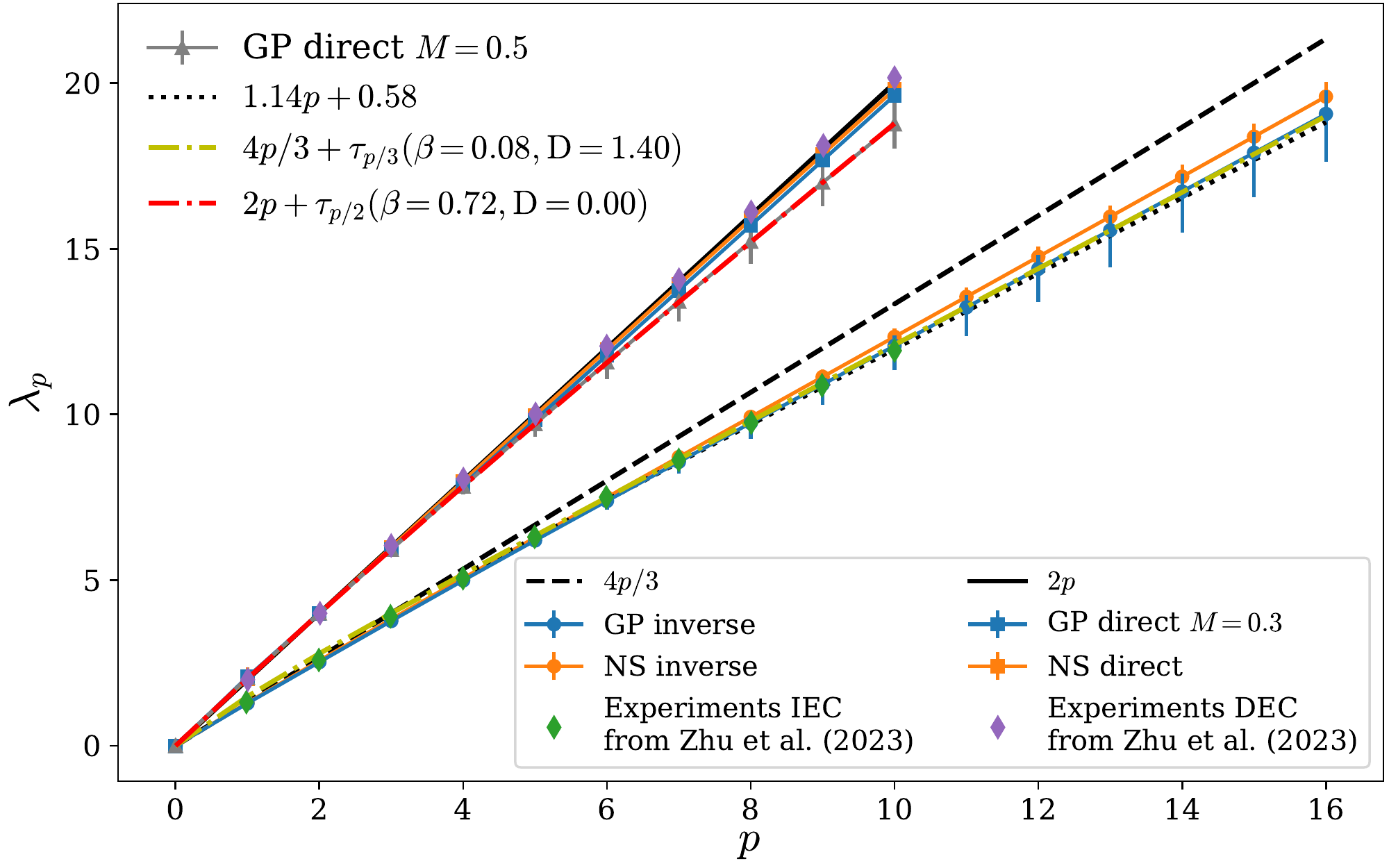}
  \caption[]{%
    Circulation scaling exponents in the inverse and direct cascade inertial ranges for both classical and quantum turbulence. Black dashed and solid lines correspond to the self-similar scaling for the inverse and direct cascade regimes, respectively. Experimental data and the dotted-line fit were extracted from \cite{Zhu2023}. Dotted-dashed lines show the fit based on the mOK62 theory of \cite{Polanco2021} using the random-$\beta$ model \cite{Benzi1984}. 
    }
    \label{fig:exponents}
\end{figure}

The circulation scaling exponents of our QT and CT simulations are presented in Fig.~\ref{fig:exponents}.
For the IEC, both systems follow the same intermittent behavior within error bars, defined as the maximum and minimum value of the local slopes in the inertial range. 
These results are consistent with recent experimental measurements in quasi-2D turbulence \cite{Zhu2023}, also reported in the figure. Moreover, the dotted line shows the monofractal fit $\lambda_p^{\mathrm{fit}} = 1.14p + 0.58$ for $p>3$, with H\"older exponent $h=1.14$ and fractal dimension $D=1.42$ proposed in \cite{Zhu2023}. Similar to 3D turbulence \cite{Muller2022a,Muller2021,Polanco2021}, CT and QT share the same statistics in 2D for the IEC.
The equivalence between CT and QT also holds for the enstrophy cascade, in which both systems display a self-similar behavior, consistent with recent experiments \cite{Zhu2023}. It is important to remark that, to recover this scaling, compressible effects in the quantum flow should be negligible. To test this idea, we repeat the previous analysis of the DEC starting from a flow with $M=0.5$. The development of quasi-shocks (Fig.\ref{fig:visualization}.d) occurs more frequently, eventually modifying the flow statistics. Figure~\ref{fig:exponents} also displays circulation scaling exponents for this run. Remarkably, low-order circulation moments still display the same scaling as the classical ones but high-orders deviate (see SM). The effect of quasi-shocks on turbulent statistics is consistent with some recent experimental measurements in compressible flows \cite{Ricard2023}.

An alternative multifractal interpretation of the intermittent behavior of velocity circulation was given in \cite{Polanco2021} by introducing a modified version of Obukhov-Kolmogorov 1962 (mOK62) theory \cite{Oboukhov1962,Kolmogorov1962}. Circulation scaling exponents are proposed to follow $\lambda_p = (h+1) p + \tau((h+1) p/4)$, where $h$ is the H\"older exponent of the velocity field, which can be related to vortex polarization \cite{Polanco2021}. For the IEC, $h=1/3$ and for the DEC $h=1$. 
The correction to the self-similar scaling $\tau(\cdot)$ is introduced through the anomalous scaling of the coarse-grained energy dissipation moments $\mean{\epsilon_r^p} \sim r^{\tau(p)}$ \cite{Dubrulle2019}. Here, we use the random-$\beta$ model of fractal dimension $D$, which reads $\tau(p) = (2-D) \left[(\beta-1)p + 1 - \beta^p\right]$, with $0<\beta<1$ a free parameter \cite{Benzi1984,Boldyrev2002}. For the inverse cascade, a best fit \footnote{We perform a least squares fit with parameters in the range $0\le\beta\le1$ and $0\le D\le 2$.} leads to $D=1.4$, in agreement with the monofractal fit of \cite{Zhu2023}. 
For the high-Mach DEC in QT, the fit yields $D=0$, suggesting isolated quantum vortices are responsible for the intermittent behavior in this regime. Note that the self-similar behavior of the CT DEC corresponds to $D=2$.

In this Letter, we reported numerical simulations of classical and quantum 2D turbulence in the direct and inverse cascade settings. Whereas several studies have been devoted to studying the inverse energy cascade in quantum turbulence \cite{Johnstone2019,Gauthier2019,Bradley2012,Reeves2013,Muller2020a}, the enstrophy cascade has only been observed using a dissipative version of the point-vortex model \cite{Reeves2017}. Here we used the Gross--Pitaevskii equation, which naturally includes vortex annihilation and interaction with sound. The observation of the DEC in GP simulations was possible thanks to the use of very high resolutions and well-controlled initial conditions that minimize acoustic emissions. 
Indeed, the enstrophy cascade only makes sense in a coarse-grained manner, as the enstrophy is not mathematically defined. Therefore, it requires a large number of vortices arranged to produce a large-scale flow. 
Moreover, we studied high-order statistics of velocity circulation in 2D classical and quantum turbulence. 
For the IEC, the intermittent behavior of CT and QT are equivalent, reminiscent of recent studies in 3D turbulence \cite{Muller2021,Polanco2021,Muller2022a}. This numerical measurement provides further support for the difference between the statistics of velocity circulation and velocity increments \cite{Iyer2019,Paret1998,Boffetta2000}.
For the DEC the equivalence between CT and QT also holds, following a self-similar scaling in both cases, provided that initial GP-flow has a low Mach number. For higher Mach flows, the equivalence only holds for low-order statistics, while the singular character of quantum vortices strongly affects high-orders. In classical fluids, shocks are smoothed out by viscous dissipation, which is very different from regularization by dispersive mechanisms in quantum flows. Naturally, it would be important to study the injection of vorticity and enhancement of circulation intermittency through strong density gradients in compressible classical fluids.
Finally, the characterization of these differences and similarities between 2D quantum and classical turbulence could be useful for the development of future theories of intermittency.

\begin{acknowledgments}
  We are grateful to Guido Boffetta for providing Navier--Stokes numerical data on the inverse energy cascade that we used in preliminary studies. 
  This work was supported by the Agence Nationale de la Recherche through the project GIANTE ANR-18-CE30-0020-01.
  GK acknowledges financial support from the Simons Foundation Collaboration grant Wave Turbulence (Award No. 651471). 
  This work was granted access to the HPC resources of CINES, IDRIS and TGCC under the allocation 2019-A0072A11003 made by GENCI.
  Computations were also carried out at the Mésocentre SIGAMM hosted at the Observatoire de la Côte d'Azur.
\end{acknowledgments}

\bibliographystyle{apsrev4-1}
\bibliography{bibliography}


\clearpage

\onecolumngrid

\section*{Supplemental Material: Exploring the Equivalence between two-dimensional Classical and Quantum Turbulence through Velocity Circulation Statistics}

\section{Generation of 2D initial condition}

We study the statistical properties of two-dimensional (2D) quantum turbulence (QT) by performing direct numerical simulations of the Gross--Pitaevksii (GP) equation 

\begin{equation}
  i \partial_t \psi = \frac{c}{\sqrt{2}\xi} \left(-\xi^2 \nabla^2 \psi + \frac{|\psi|^2}{n_0} \psi - \psi\right)
  \label{eq:gp_sm}
\end{equation}

\noindent with $\psi$ the condensate wave function, $c$ the speed of sound, $\xi$ the healing length and $n_0$ the ground state particles density. We set $n_0=1$ and the healing length so that $\xi k_{\mathrm{max}} > 2$ to solve well the density profile of quantum vortices. 

The total energy in the GP equation is conserved, with the incompressible energy being irreversibly transferred into sound. Therefore, when it comes to studying the quantum vortex dynamics, GP simulations can be seen as decaying quantum turbulent runs. Statistical properties of the flow are typically studied during a transient period in which turbulence is strongest. 
The implementation of an external forcing would also excite phonons that accelerate the process of vortex annihilation and contaminate the dynamics of quantum vortices. 

To generate the 2D initial condition with the initial energy concentrated between some target wave numbers, we build a velocity field in the Clebsch representation $\vvec = \lambda \gradient \mu - \gradient \chi$ \cite{Nore1997, Muller2020a}, where the Clebsch potential are a linear combination of modes

\begin{align}
  \lambda &= \frac{1}{2 k_\lambda} \sum_{k_i=1}^{2k_\lambda} \cos \left\{ x \left[ k_\lambda \cos \left(\frac{\pi k_i}{2 k_\lambda} \right) \right]  + y \left[ k_\lambda \sin \left(\frac{\pi k_i}{2 k_\lambda}\right) \right] + \phi_{k_i} \right\} \\
  \mu &= \frac{1}{2 k_\mu} \sum_{k_i=1}^{2k_\mu} \cos \left\{ x \left[ k_\mu \cos \left(\frac{\pi k_i}{2 k_\mu}\right) \right] + y \left[ k_\mu \sin \left(\frac{\pi k_i}{2 k_\mu}\right) \right] + \varphi_{k_i} \right\} 
\end{align}

\noindent where $k_\lambda$ and $k_\mu$ set the maximum number of the superposition of modes for each Clebsch potential, $\phi_{k_i}$ and $\varphi_{k_i}$ are random phase, and the square brackets $\left[ . \right]$ indicate the integer part of the argument to satisfy periodicity. The third Clebsch potential $\chi$ is defined so that the initial velocity field is incompressible $\gradient \cdot \vvec = 0$. Different values of $k_\lambda$ and $k_\mu$ will lead to different initial conditions with energy concentrated at different scales. The vorticity field is defined as $\vortvec = \gradient \times \vvec = \gradient \lambda \times \gradient \mu$. 
With these potentials, we can also construct a wave function $\psi$ such that $\psi(x,y) = \psi_e(\lambda(x,y), \mu(x,y))$ where the $\psi_e$ has zeros at the zeros of the Clebsch potentials, corresponding to quantum vortices with a non-zero vorticity field. 
The wave function is then constructed as $\psi_e(\lambda, \mu) = (\lambda + i\mu) \tanh \left[ \sqrt{\lambda^2 + \mu^2} / (\sqrt{2}\xi) \right] / \sqrt{\lambda^2 + \mu^2}$. The choice of a hyperbolic tangent is done to give a good approximation of the quantum vortex density profiles. In practice, we can increase the number of vortices by multiplying several of these wave functions, for example, $\psi(x, y) = \Pi_{i=-1}^1 \psi_e(\lambda-i, \mu-i)$. 

As a final step, we want the wave function to follow the velocity field generated by the Clebsch potentials, with a minimal amount of acoustic emission. In particular, we generate initial flows with a Mach number $M=U/c\leq 0.3$ to neglect compressible effects. For this purpose, we first evolve the wave function using the advective real Ginzburg-Landau (ARGL) equation

\begin{equation}
  \partial_t \psi = -\frac{c}{\sqrt{2}\xi} \left(-\xi^2 \nabla^2 \psi + \frac{|\psi|^2}{n_0} \psi - \psi\right) - i \vvec \cdot \gradient \psi - \frac{(\vvec)^2}{2 \sqrt{2} c\xi} \psi.
  \label{eq:argl}
\end{equation}

\noindent Once the evolution of the ARGL equation is well converged, we evolve the system using the GP Eq.~\eqref{eq:gp_sm}. 

\section{Parameters of the numerical simulations}

To study the properties of 2D classical and quantum turbulence, we solve the GP equation incompressible Navier--Stokes (NS) equation, which in terms of the vorticity $\omega$ is written as 

\begin{equation}
  \partial_t \omega + \left\{\omega, \phi\right\} = \nu \nabla^2 \omega - \alpha \omega + f
  \label{eq:ns_sm}
\end{equation}
with $\phi$ the stream function such that the velocity field is $(u, v) = (\partial_y \phi, -\partial_x \phi)$, the Poisson brackets are defined as $\left\{\omega, \phi\right\} = \partial_x \omega \partial_y \phi - \partial_y \omega \partial_x \phi$, $\nu$ is the kinematic viscosity, $\alpha$ is a linear friction preventing the formation of a large-scale condensate, and $f$ an external forcing. 
We solve these equations using a standard pseudospectral method in a bi-periodic domain and a Runge--Kutta method for time stepping of order 2 for NS and order 4 for GP. 
We perform a total of four numerical simulations, two to study the inverse energy cascade (IEC), and another two for the direct enstrophy cascade (DEC). Out of these two, in one we solve the NS equation and in the other the GP equation. 

For the NS IEC simulation, we force at small scales using a Gaussian forcing centered at $k_f = 637$ of width $\Delta k = 425$ with $N=6144$ linear collocation points, $\nu=8\times 10^{-6}$ and $\alpha=3.28 \times 10^{-2}$, while for the DEC we use a random forcing at $k_f=1$ and $k_f=2$ with $N=6144$, $\nu=5\times 10^{-7}$ and $\alpha=0$. 
Both of these regimes are studied in a stationary state and we use several hundred velocity fields to increase the amount of statistics.
For the GP simulations, we use $N=8192$ linear collocation points and set the healing length $\xi=1.5 \Delta x$, $n_0=1$ and the speed of sound $c=1$, with $\Delta x = L/N$ and $L=2\pi$ the size of the domain. 
To study the IEC, we generate an initial condition as described in the previous section, with wave numbers in a band around $k_{\mathrm{IC}}=30$, and an initial Mach number $M=\vrms/c=0.3$. For the DEC we initialize a flow with wave numbers between $k_{\mathrm{IC}}=1$ and $k_{\mathrm{IC}}=2$, and $M=0.29$. 
As these simulations are not in a stationary state, we only study small time intervals in which turbulence is stronger. To increase the amount of statistics, we perform an ensemble of four and five numerical simulations for the inverse and direct cascades, respectively. These runs are performed using different random phases for the initial conditions and are statistically equivalent. To study compressible effects, we compare some of the results with an ensemble of seven runs with initial Mach $M=0.5$.

\section{Energy and enstrophy evolution in Gross--Pitaevskii}

\begin{figure}
  \centering
  \tikzmark{a}\includegraphics[width=.48\textwidth]{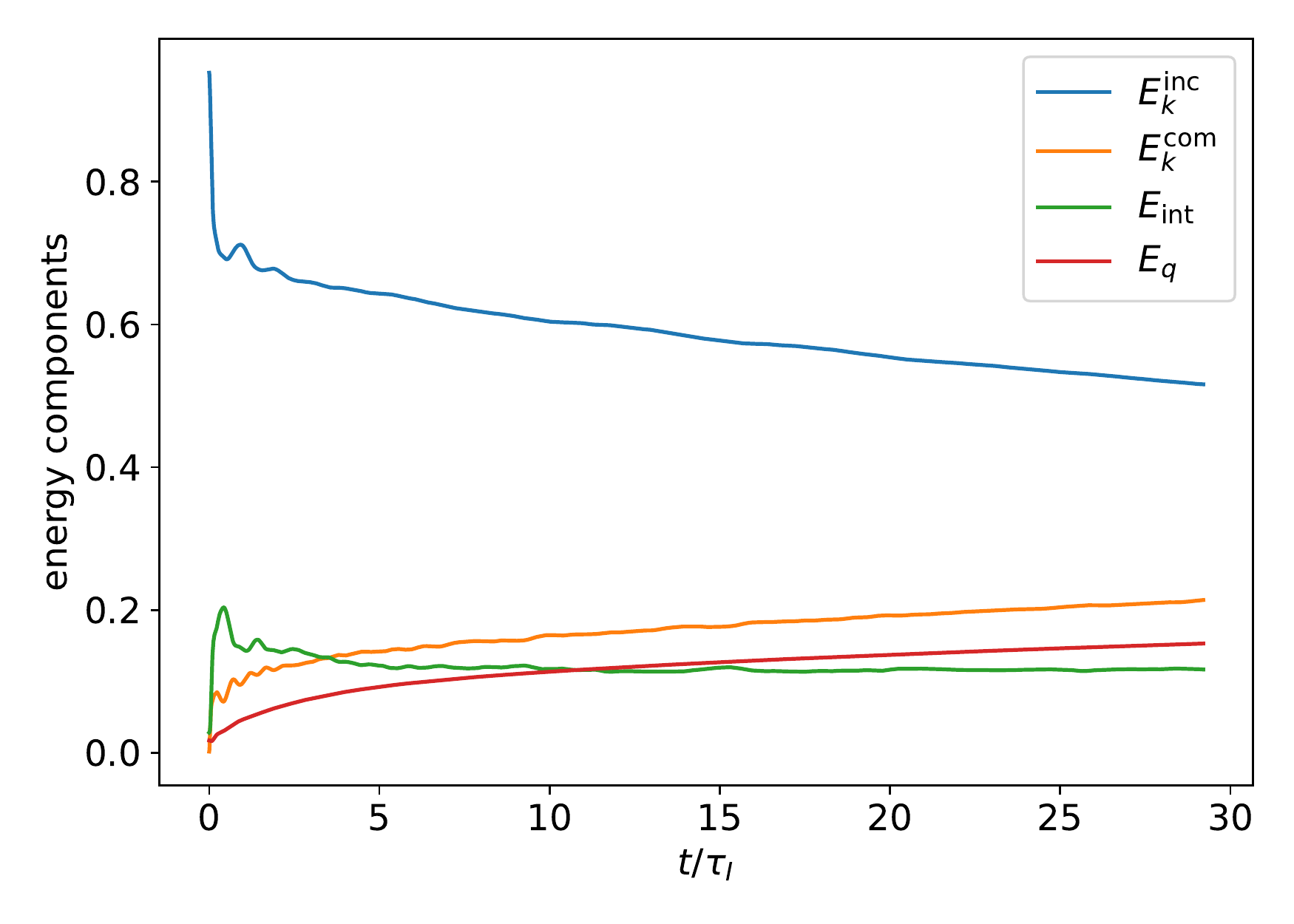}
  \tikzmark{b}\includegraphics[width=.48\textwidth]{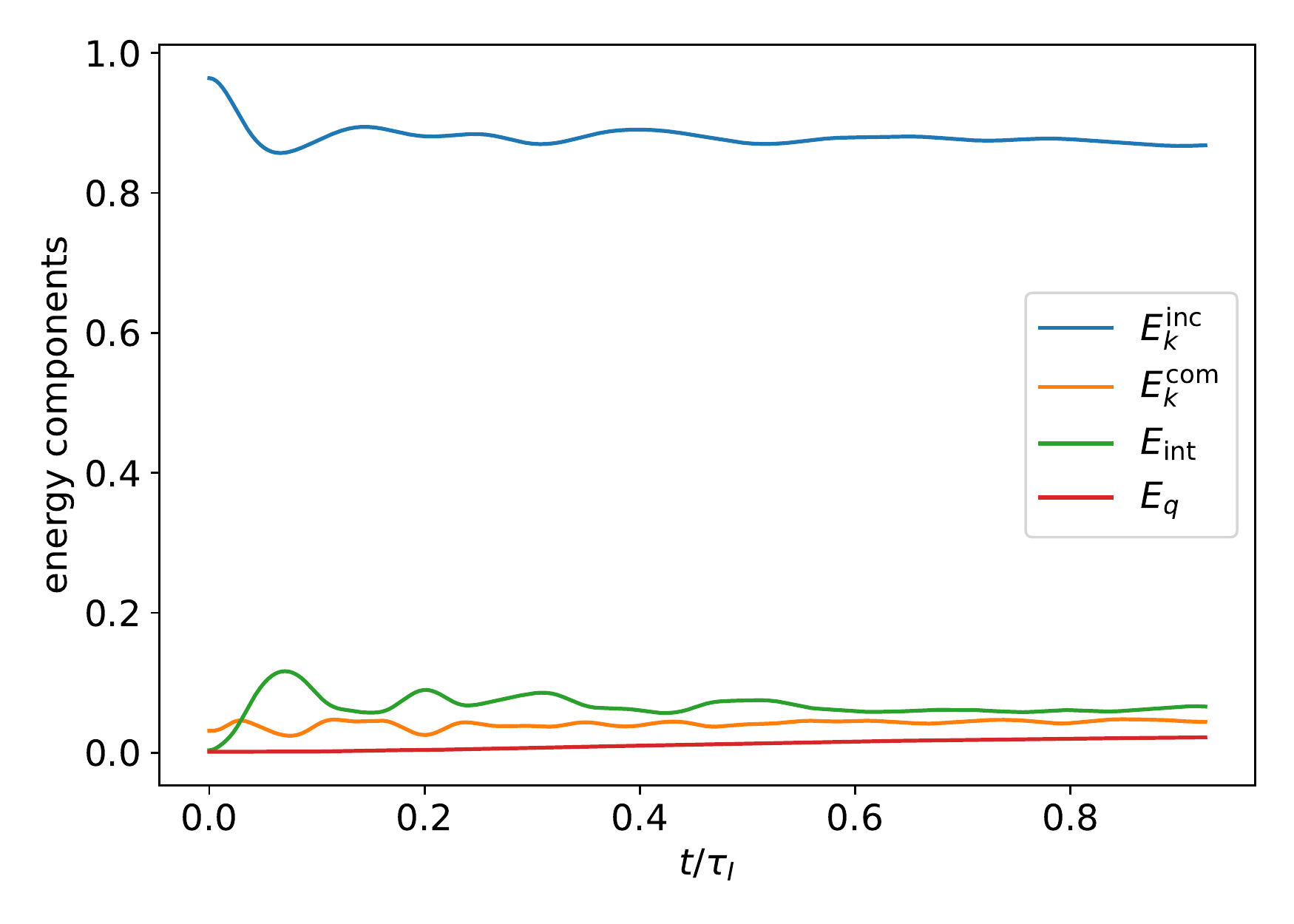}
  \caption[]{%
        Evolution of the incompressible, compressible, internal and quantum energy components in (a) the inverse energy cascade and (b) in the direct enstrophy cascade simulations with Mach number $M=0.3$. }
    \label{fig:energy_evolution}
\end{figure}
\InsertLabelsss{8ex}{3ex}

As the simulations of the GP equation are performed without any external forcing, there are different regimes that can be found in a 2D quantum turbulent flow \cite{Shukla2013}. Among these, there is first a regime in which the initial conditions of the system dominate, a turbulent regime, a decaying one and eventually for long times the system can thermalize due to finite size effects. 
In our case, we will study the statistical properties of the flow within a time window in which turbulence is strongest and when the system displays good Kolmogorov-like scaling properties. This is a standard method used in classical and quantum decaying turbulence \cite{Mininni2013,ClarkdiLeoni2017,Nore1997,Barenghi2014a}. 
Figure \ref{fig:energy_evolution} shows the evolution of the different energy components of the GP-inv and GP-dir-M03 simulations. We made sure that in the whole turbulent regime, the incompressible kinetic energy (the one displaying Kolmogorov-like properties and the relevant one for circulation) is larger than the other components at all times. Time is normalized by the turnover time $\tau_I = v_{\mathrm{rms}} / L_0$ with $L_0$ the initial integral length scale. 

\begin{figure}
  \centering
  \tikzmark{a}\includegraphics[width=.48\textwidth]{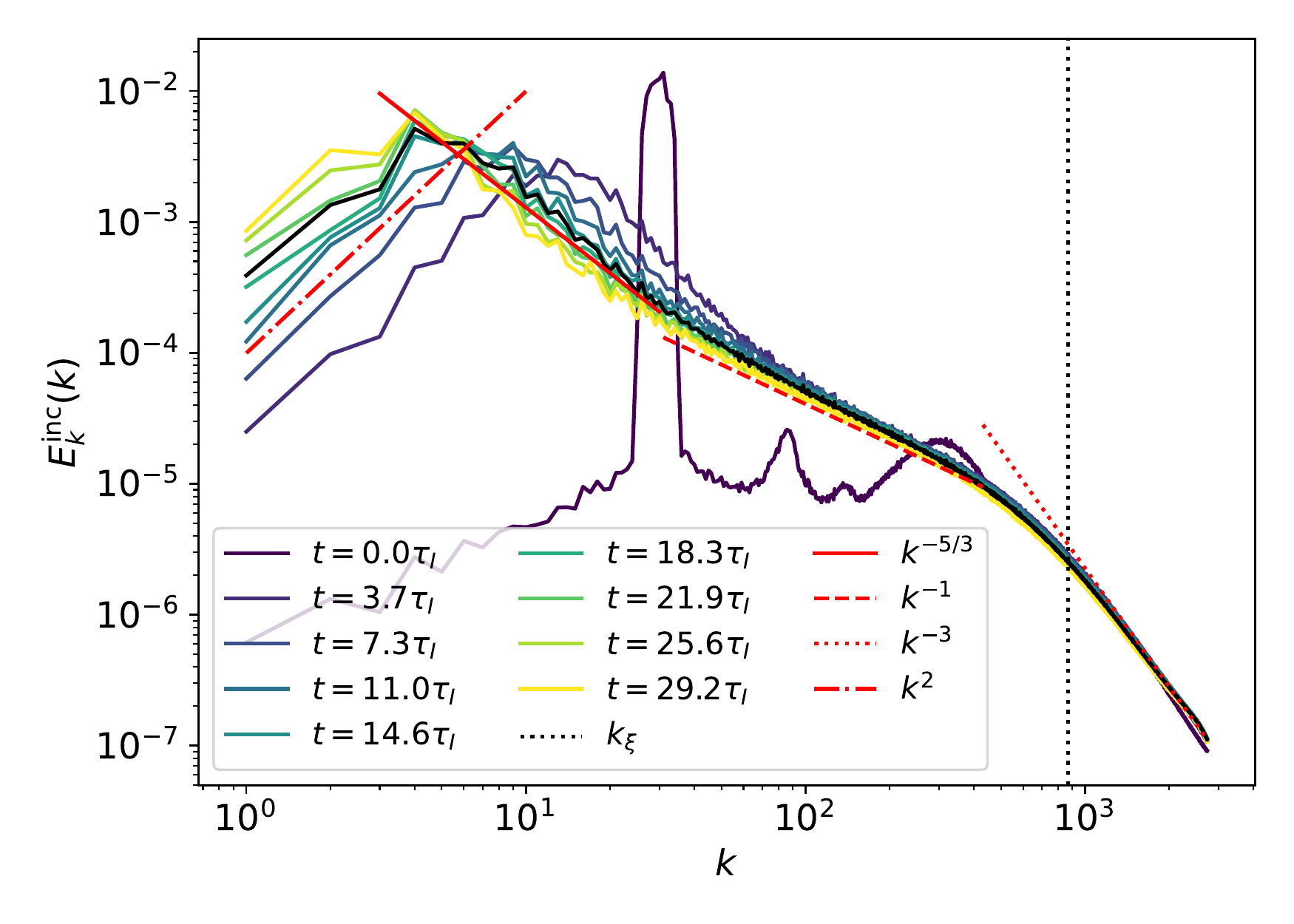}
  \tikzmark{b}\includegraphics[width=.48\textwidth]{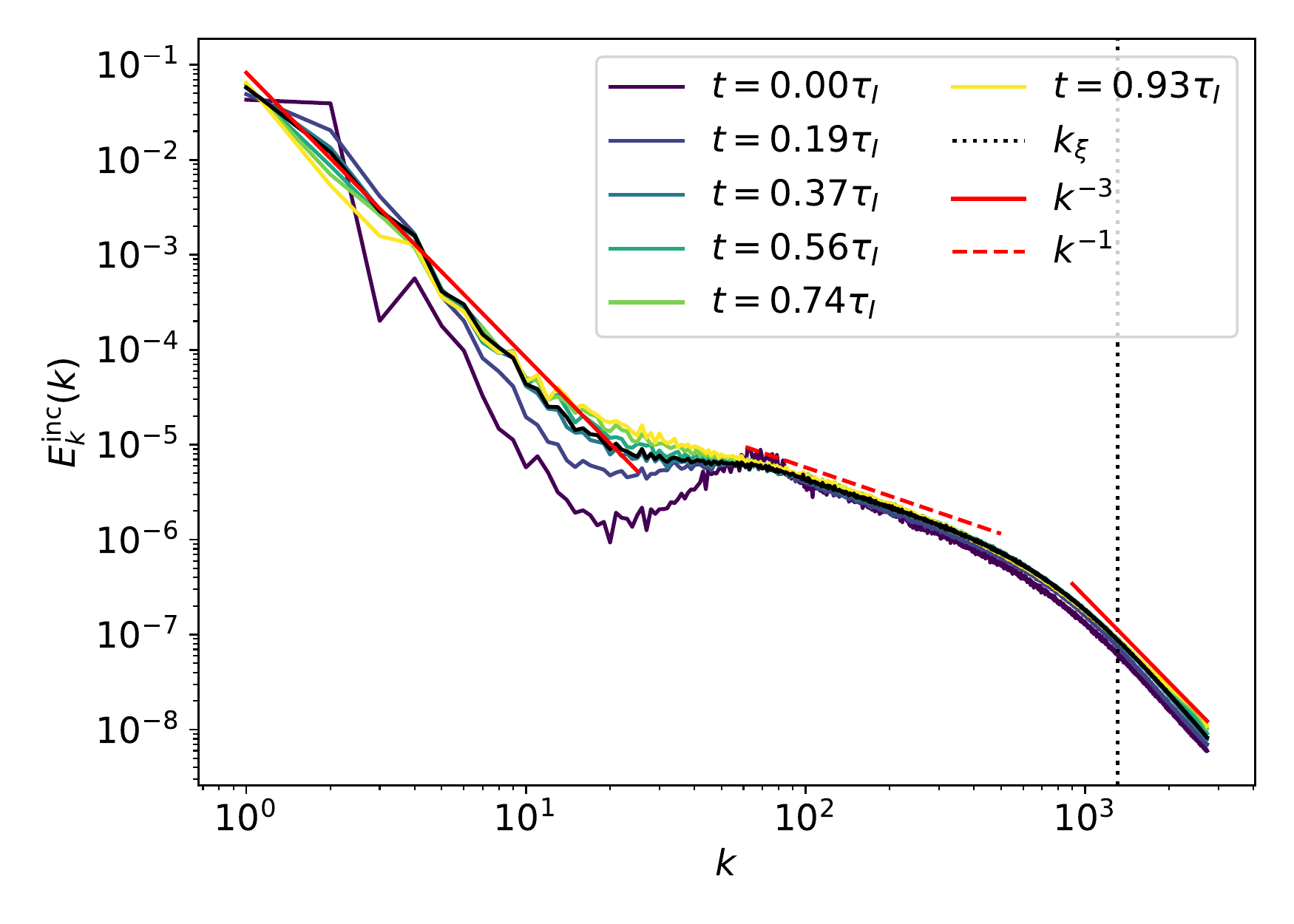}
  \tikzmark{c}\includegraphics[width=.48\textwidth]{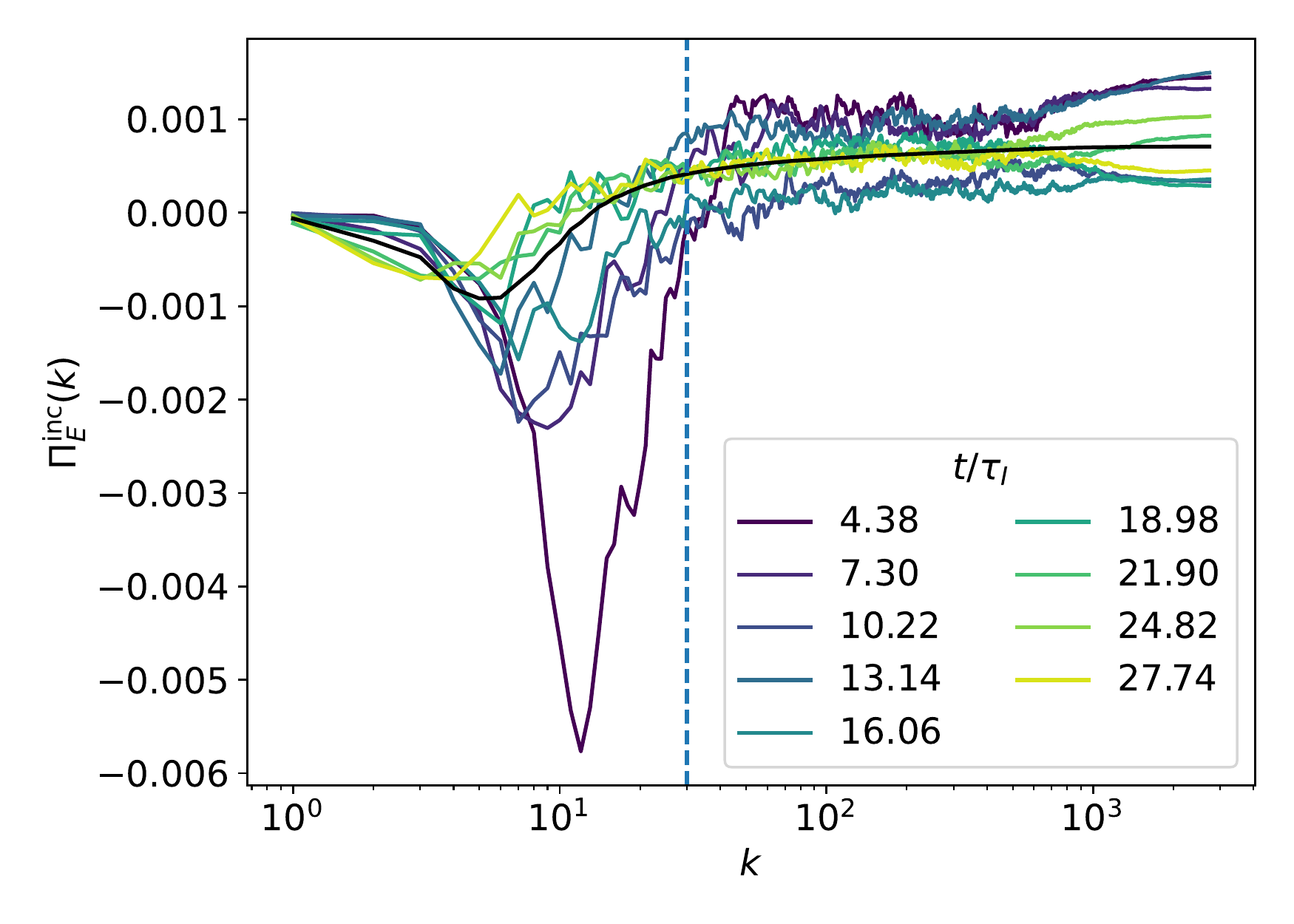}
  \tikzmark{d}\includegraphics[width=.48\textwidth]{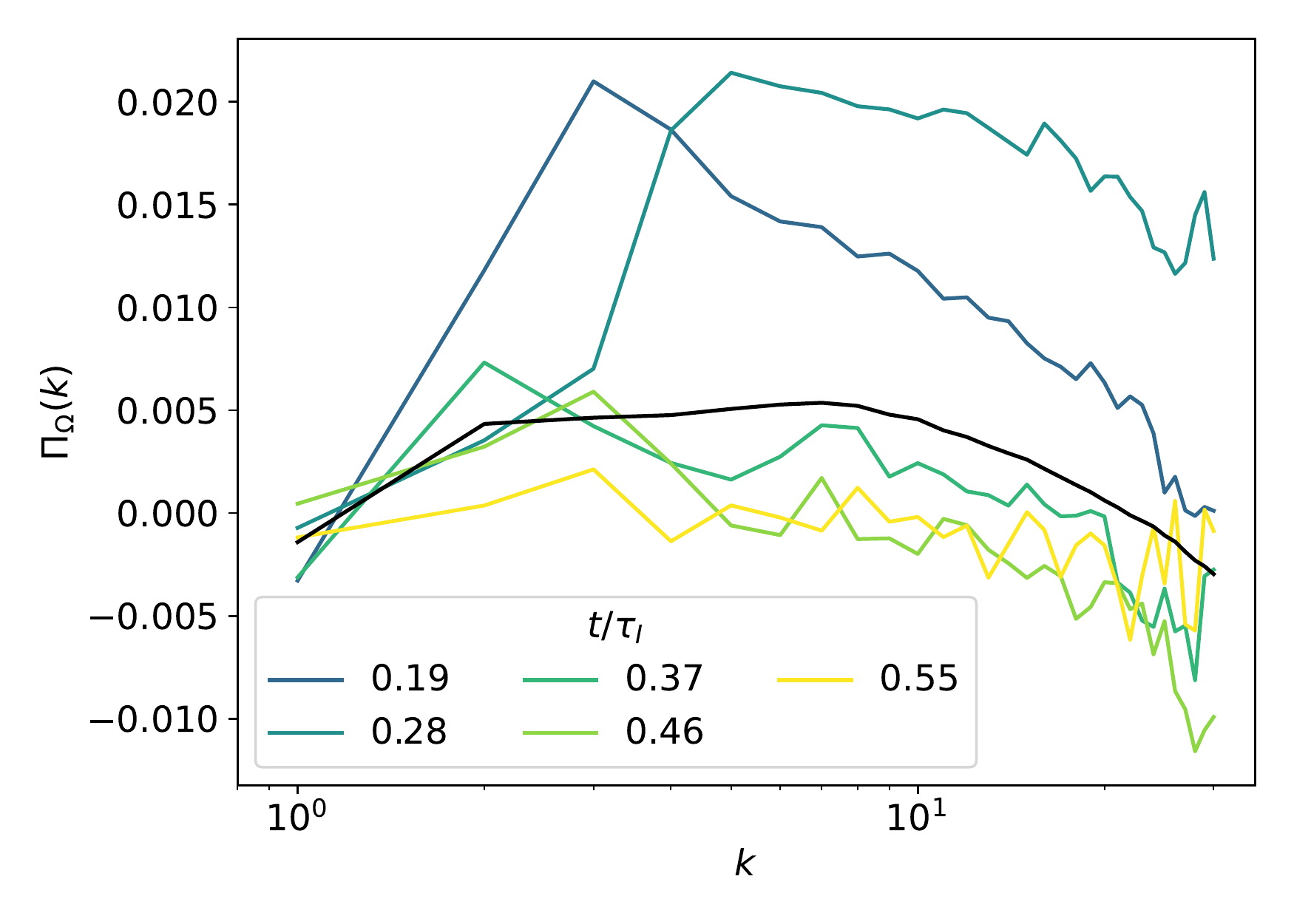}
  \caption[]{%
  Evolution of the (a)-(b) incompressible energy spectra, (c) incompressible energy flux and (d) enstrophy flux. The left panels correspond to the inverse energy cascade and the right panels to the direct enstrophy cascade simulations. Black solid lines correspond to temporal averages. }
    \label{fig:spectra_evolution}
\end{figure}
\InsertLabels{12ex}{2ex}

Figure \ref{fig:spectra_evolution} (a) and (b) show the evolution of the energy spectra in the GP-inv and GP-dir-M03 simulations, respectively. In the former case, we excite the flow at a range of wave numbers centered at $k_0\approx30$. As it evolves, the peak of the spectra moves to larger scales, a signature of an inverse energy cascade process \cite{Muller2020a}. Among the different scales and different times, we observe the development of different scaling laws. 
At scales larger than the peak, there is a $k^2$ scaling law that is larger than the expected scaling of energy equipartition $k^1$. The origin of this scaling regime is not completely clear and further studies are left for future work. 
Between the spectral peak and $k_0$, there is a range of scales in which the fluid develops Kolmogorov-like scaling properties, satisfying a $-5/3$ scaling law. Scales smaller than $k_0$ also correspond to scales smaller than the intervortex distance, as in the case of IEC simulations $k_0 \approx k_\ell$, so we observe the development of a $k^{-1}$ energy spectrum given by individual vortex lines. Finally, at scales smaller than the healing length, dispersive effects dominate and we enter a $k^{-3}$ scaling regime \cite{Krstulovic2010}. For the DEC, initially, the flow at large scales develops a $k^{-4}$ scaling property and as it evolves it approaches the $k^{-3}$ scaling law predicted by Kraichnan, neglecting logarithmic corrections \cite{Kraichnan1967}. At scales smaller than the intervortex distance, the scaling laws of individual vortex lines are again recovered, as in the IEC cascade. 

The individual observation of the celebrated $k^{-5/3}$ and $k^{-3}$ scaling laws in the incompressible energy spectra are not enough to establish the presence of an inverse and direct cascade, respectively. These scaling laws should also be accompanied by constant energy and enstrophy fluxes, as shown in Fig.~\ref{fig:spectra_evolution} (c)-(d). We define the incompressible energy and enstrophy fluxes as \cite{Muller2020a}

\begin{gather}
  \Pi_E^{\mathrm{inc}}(k) = - \sum_{\tilde{k}=1}^{\tilde{k}=k} \frac{\dd E_k^{\mathrm{inc}}(\tilde{k})}{\dd t} \label{eq:Eflux} \\ 
  \Pi_\Omega(k) = - \sum_{\tilde{k}=1}^{\tilde{k}=k} \tilde{k}^2 \frac{\dd E_k^{\mathrm{inc}}(\tilde{k})}{\dd t}.
  \label{eq:ens_flux}
\end{gather}
Indeed, in the IEC simulations for some intermediate times, there is a range of scales where $\Pi_E^{\mathrm{inc}} < 0$ and is almost constant. Note that $\Pi_E^{\mathrm{inc}}(k_{\mathrm{max}}) > 0$ meaning that incompressible energy is transferred to the other components at small scales. For the DEC cascade, we observe the signature of an enstrophy cascade as $\Pi_\Omega>0$ and almost constant for $k_0<k<k_\ell$. We show only the enstrophy flux for scales larger than the intervortex distance as it is not properly defined at smaller scales due to the singular nature of quantum vortices. 
Black solid lines show temporal averages of the different quantities. 

\begin{figure}
  \centering
  \includegraphics[width=.9\textwidth]{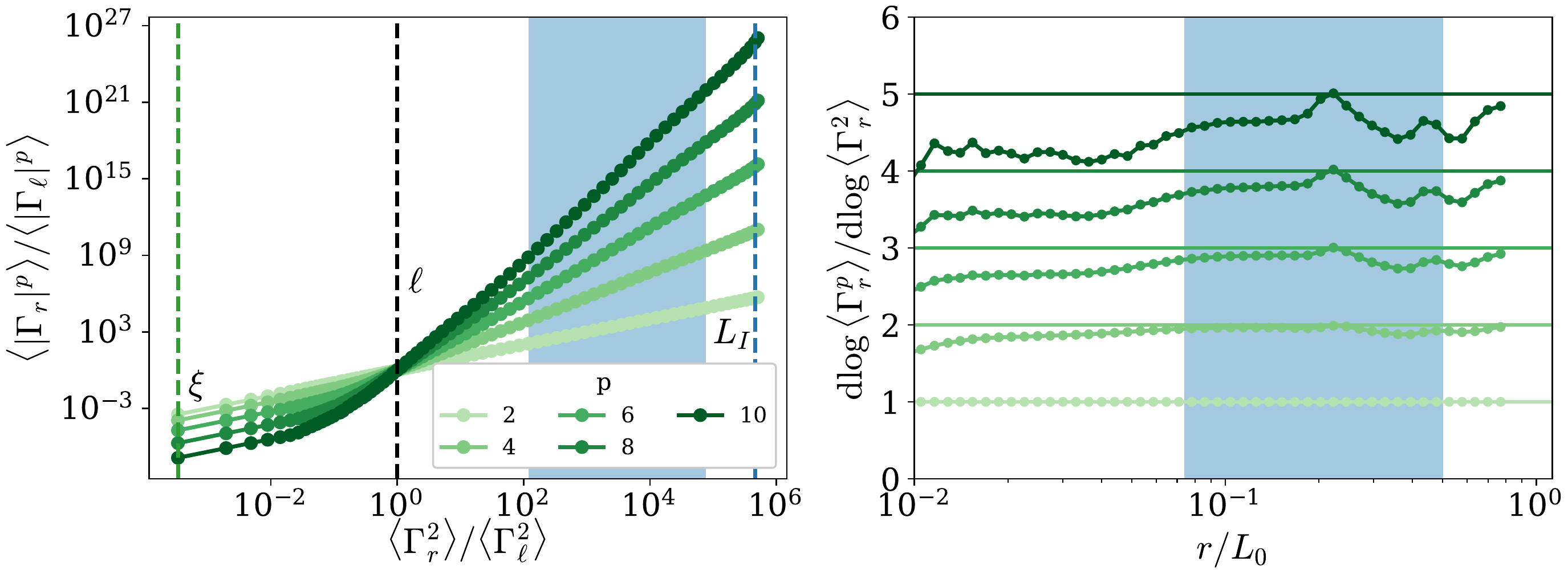}
  \caption[]{%
    Circulation moments in numerical simulation of GP-dir-M05 run up to $p=10$. The blue-shaded area corresponds to the inertial range in which the scaling exponents shown in Fig.~4 of the main text are computed. }
    \label{fig:circulation_moments_gp_dir_M05}
\end{figure}

High-order moments of circulation in the GP-inv and GP-dir-M03 are shown in the main text of this manuscript. To compare, we include in Fig.~\ref{fig:circulation_moments_gp_dir_M05} the circulation moments of run GP-dir-M05, in which compressible effects can not be neglected. In this case, we observe some deviations from the self-similar prediction for the direct cascade of $\lambdadir=2p$, suggesting that the development of quasi-shocks enhances the intermittent behavior of the flow.  

\section{Circulation moments in Navier--Stokes}

\begin{figure}
  \centering
  \includegraphics[width=.9\textwidth]{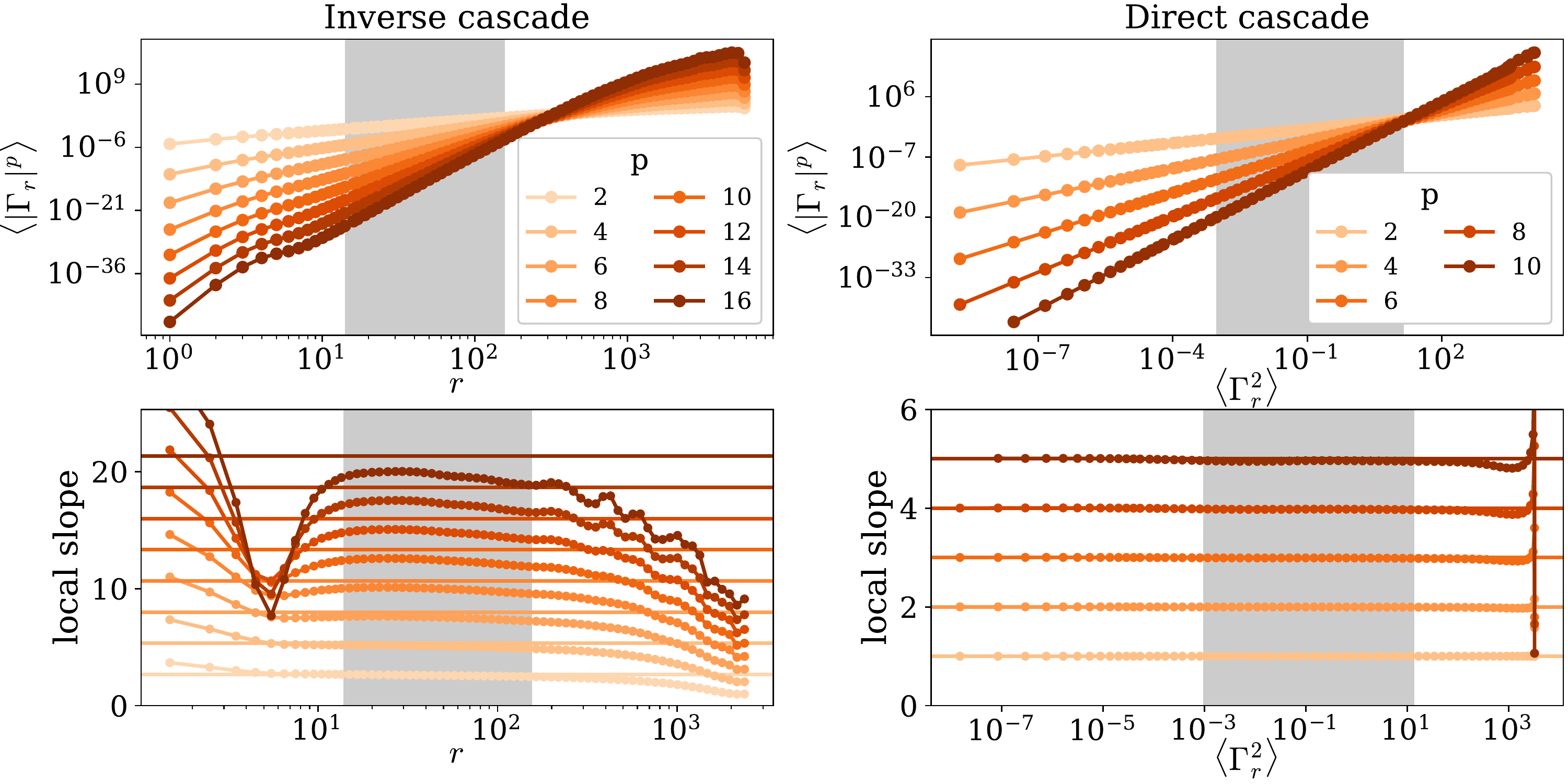}
  \caption[]{%
    Circulation moments in numerical simulation of the 2D Navier--Stokes equations for (left panel) the inverse energy cascade up to order $p=16$ and (right panel) the direct enstrophy cascade up to $p=10$. For the DEC, we use the extended-self-similar (ESS) hypothesis and plot as a function the moments of $\langle \Gamma_r^2 \rangle$. The bottom panels display the local slopes defined as the logarithmic derivatives of the moments of order $p$.}
    \label{fig:moments_ns}
\end{figure}

In both the DEC and IEC regimes of 2D classical turbulence, we study the statistics of circulation and characterize its intermittent behavior. 
Figure~\ref{fig:moments_ns} shows the circulation moments $\mean{\Gamma_r^p}$ up to $p=16$ for the IEC and $p=10$ for the DEC. The insets show the local slopes, defined as the logarithmic derivatives $\dd \log \mean{\Gamma^p} / \dd \log r$. For the IEC, the local slopes display a deviation from the self-similar prediction $\lambdainv_p = 4p/3$ within the inertial range of scales indicated as the gray shaded area. In the DEC regime, we use the extended-self-similar (ESS) method with respect to $\mean{\Gamma^2_r}$ and obtain that the scaling exponents become self-similar following the prediction $\lambdadir_p=2p$. 

\begin{figure}
  \centering
  \includegraphics[width=.9\textwidth]{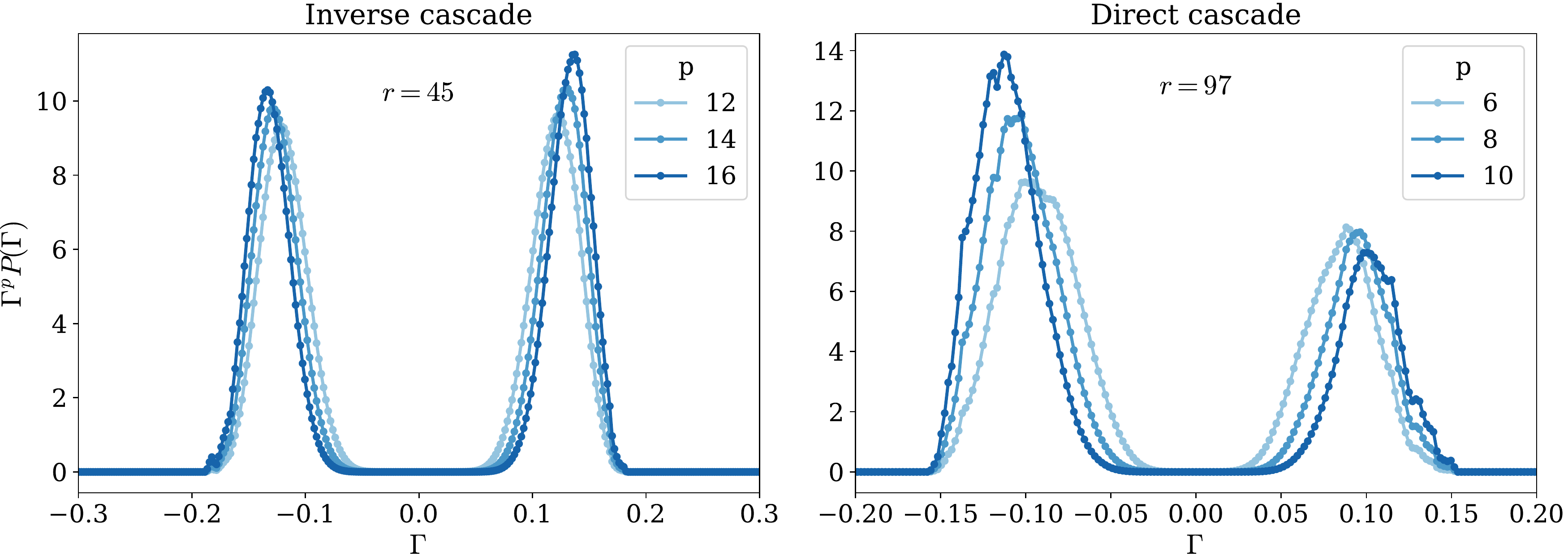}
  \includegraphics[width=.9\textwidth]{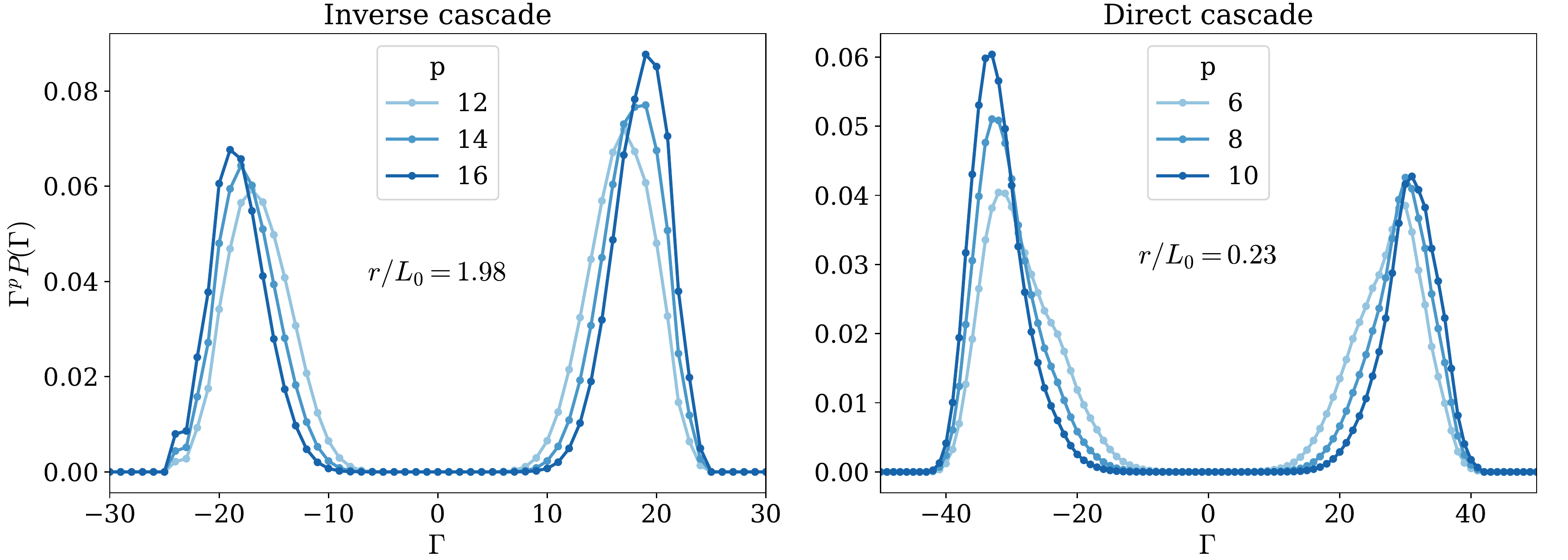}
  \caption[]{%
    Circulation integrands for (top panel) 2D classical turbulence and (bottom panel) quantum turbulence in the inverse and direct cascade regimes for a length scale within the inertial range. }
    \label{fig:integrands}
\end{figure}

To check for the convergence of these moments, we make sure that the integrands $\Gamma^p P(\Gamma)$ go to zero when $\Gamma \rightarrow \pm \infty$. If this condition is not satisfied, it means that there is not enough statistics to study high-order moments.
These integrands are shown in Fig.~\ref{fig:integrands}. Indeed, we observe that for the studied order moments the probability distribution functions (PDFs) in the IEC and DEC regimes are well resolved for length scales within the inertial range. 
The bottom panels show the circulation integrands for the Gross--Pitaevskii GP-inv and GP-dir-M03 simulations, again displaying a good convergence. 

\section{Movies}

We include the time evolution of the 2D GP dynamics in four different numerical simulations. 
The movies are scatter plots of the positions of positive (red) and negative (blue) individual vortices extracted from the wave function using a circulation-based algorithm. For some frames, this algorithm fails to find some of the vortices. Three movies correspond to one of the individual runs from the ensemble called the GP-dir-M03, GP-dir-M05 and GP-inv in the main manuscript. We provide a fourth movie that shows a longer evolution of the flow in the inverse cascade. 

The movies are: 

- \textit{gp2d\_8192\_k1\_M03.mp4}: Evolution of the GP-dir-M03 run. Initial condition generated at $k_0=1$ with a Mach number $M=0.3$. 

- \textit{gp2d\_8192\_k1\_M05.mp4}: Evolution of the GP-dir-M05 run. Initial condition generated at $k_0=1$ with a Mach number $M=0.5$. Compressible effects become more evident.

- \textit{gp2d\_8192\_k30.mp4}: Evolution of the GP-inv run. Initial condition generated at $k_0=30$. 

- \textit{gp2d\_8192\_k20.mp4}: Extra movie not analyzed in the manuscript. Initial condition generated at $k_0=20$. 

\end{document}